\documentclass[journal]{IEEEtran}
\usepackage{url} 
\usepackage{cite}
\usepackage{amsmath,amssymb,amsfonts}
\usepackage{algorithmic}
\usepackage{graphicx}
\usepackage{textcomp}
\usepackage{xcolor}
\usepackage{threeparttable}
\usepackage{wrapfig}

\begin{document}
\title{Towards Ultrasensitive SQUIDs Based on Submicrometer-Sized Josephson Junctions}

\author{Jan-Hendrik Storm, Oliver Kieler, and Rainer K\"orber%
\thanks{Manuscript received December 19, 2019; revised March 31, 2020; accepted April 17, 2020. This work was supported in part by the European Union's Horizon 2020 research and innovation program under Grant 686865, by the DFG under Grant KO 5321/1-1 and Grant KI 698/3-2. This paper was recommended by Associate Editor H. Rogalla. \textit{(Corresponding author: Rainer K\"orber.)}

J.-H. Storm was with Physikalisch-Technische Bundesanstalt, 10587 Berlin, Germany.

O. Kieler is with Physikalisch-Technische Bundesanstalt, 38116 Braunschweig, Germany.

R. K\"orber is with Physikalisch-Technische Bundesanstalt, 10587 Berlin, Germany. (e-mail: rainer.koerber@ptb.de)

Color versions of one or more of the figures in this paper are available online at http://ieeexplore.org 

Digital Object Identifier 10.1109/TASC.2020.2989630}}

\markboth{IEEE TRANSACTIONS ON APPLIED SUPERCONDUCTIVITY}%
{Shell \MakeLowercase{\textit{et al.}}: Submicrometer-sized Josephson Junctions for ultra-sensitive SQUIDs}
\maketitle

\begin{abstract}
We recently demonstrated a 1$^{\textrm{st}}$-order axial gradiometer SQUID system, which is operated in a liquid He dewar with negligible noise contribution. The achieved close to SQUID-limited measured coupled energy sensitivity $\varepsilon_{c}$ of $\sim 30\,h$ corresponds to a white field noise below 180~aT~Hz${^{-1/2}}$. In order to further improve the SQUID noise performance, the junction capacitance was reduced by decreasing its lateral size from $2.5~\mu$m to below $1~\mu$m. This was realized by extending the fabrication process for submicrometer-sized Josephson Junctions based on the HfTi self-shunted junction technology to an SIS process with AlO$_{\textrm{x}}$ as the insulating layer. We achieved energy sensitivities of 4.7$\,h$ and 20$\,h$ at 4.2~K for uncoupled and coupled SQUIDs, respectively. We also investigated the temperature dependence of the noise of the uncoupled SQUIDs and reached an energy sensitivity of 0.65$\,h$ in the white noise regime at 400~mK.
\end{abstract}


\section{Introduction}

Superconducting quantum interference devices (SQUIDs) are the most sensitive detector for magnetic flux, and state-of-the-art PTB current sensor SQUIDs reach a coupled energy sensitivity of about 30$\,h$ when operated at 4.2~K ($h$ is Planck's constant). For the detection of weak magnetic signals from room temperature samples, e.g., encountered in biomagnetism, SQUIDs are typically coupled to a superconducting pick-up coil and operated in a glass fiber liquid He dewar. In our latest system, the thermal noise from the superinsulation and thermal shields could be avoided enabling a close to SQUID-limited  white magnetic field noise $S_{B}^{1/2}$ below 180~aT~Hz${^{-1/2}}$ for a 45~mm diameter gradiometric pick-up coil~\cite{Storm2017, Storm2019}. Hence, improvements in SQUID performance would not only be beneficial for biomagnetism, but also for other applications where SQUID noise is the limiting factor.

For an uncoupled dc~SQUID of inductance $L_{\textrm{SQ}}$, critical current $I_{\textrm{c}}$, shunt resistance $R$ and junction capacitance $C$, the design parameters ${\beta_{c}=2\pi I_{\textrm{c}} R^{2}C/\Phi_{0}}$ and $\beta_{L}=2 L_{\textrm{SQ}}I_{\textrm{c}}/\Phi_{0}$ are chosen close to 1 for optimal noise performance. In this case, numerical simulations yield for the energy sensitivity per unit bandwidth $\varepsilon\approx 16 k_{\textrm{B}}T(L_{\textrm{SQ}}C)^{1/2}$, where $k_{\textrm{B}}$ is the Boltzmann constant and $T$ the temperature~\cite{Clarke2004}. 

The energy sensitivity of the uncoupled SQUID is determined experimentally by ${\varepsilon=S_{\Phi}/(2L_{\textrm{SQ}})}$, where $S_{\Phi}$ is the measured flux noise power density. For a SQUID with integrated input coil (current sensor SQUID), the coupled energy sensitivity ${\varepsilon_{c}=\varepsilon/k^{2}}$ is referred to the input coil (with inductance $L_{\textrm{i}}$) via the coupling coefficient $k=M_{\textrm{i}}/(L_{\textrm{SQ}}L_{\textrm{i}})^{1/2}$. Here, $M_{\textrm{i}}$ is the mutual inductance between the input coil and the SQUID loop. The equivalent field noise can be obtained from $S_{B}^{1/2}=S_{\Phi}^{1/2}L_{\textrm{tot}}/(M_{\textrm{i}}A_{\textrm{p}})=(2\varepsilon_{c}/L_{\textrm{i}})^{1/2}L_{\textrm{tot}}/A_{\textrm{p}}$, where $L_{\textrm{tot}}$ is the total inductance of the input circuit and $A_{\textrm{p}}$ the field sensitive area of the pick-up loop.

The aforementioned simulations show that an improvement in the energy resolution $\varepsilon$ is possible by the following:  

\begin{enumerate}
	\item lowering the SQUID inductance $L_{\textrm{SQ}}$ by decreasing the size of the SQUID loop;
	\item reducing the Josephson junction (JJ) capacitance $C$ by decreasing the junction area;
	\item  cooling down the SQUID device to reduce thermal noise in the shunt resistors.
\end{enumerate}

While approach 1) has been implemented in nano-SQUIDs, it is impractical for current sensor SQUIDs with high inductance pick-up coils (${\sim\mu}$H) as coupling to the small SQUID loop becomes excessively difficult.

Consequently, in this work, we present our development of dc~SQUIDs based on submicrometer-sized Josephson Junctions to reduce the JJ capacitance and thereby increase sensitivity. We also present the noise performance for temperatures as low as 400~mK. This is important for applications where the SQUIDs are cooled to below 4.2~K.

\section{Submicrometer-sized Josephson Junctions}

\subsection{Junction Technology}

\begin{figure}[t]
\centerline{\includegraphics[width=.90\columnwidth]{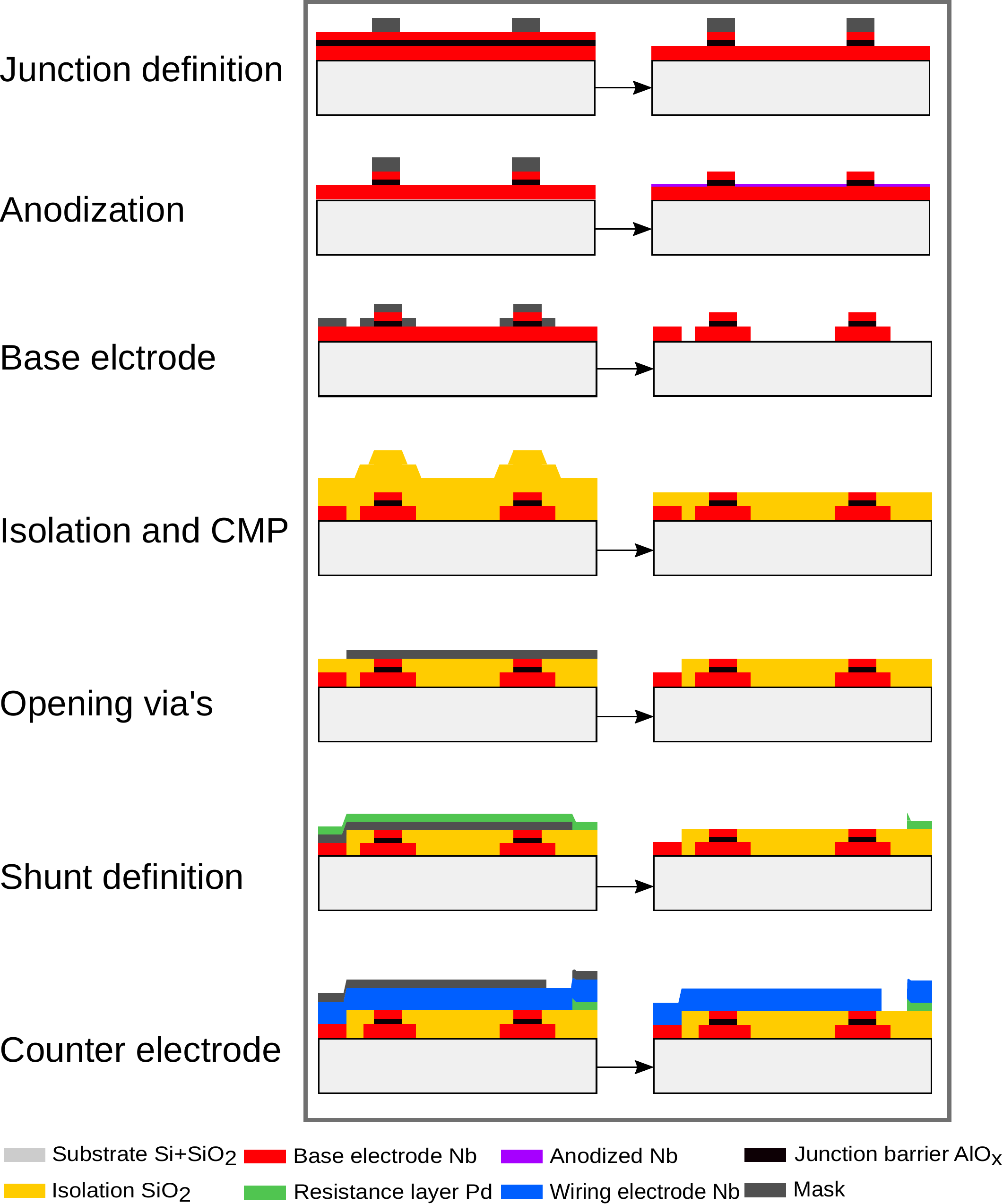}}
\caption{Technology for submicrometer-sized JJs.}
\label{fig:figure1}
\end{figure}

SQUIDs based on cross-type submicron-sized JJs have been reported in the literature~\cite{Schmelz2017,Luomahaara2018}. We chose a fabrication process for the submicrometer-sized JJs based on the established HfTi self-shunted junction technology developed at PTB for JJs arrays~\cite{Hagedorn2006} and nano-SQUIDs~\cite{Bechstein2017}. The technology has been extended to a superconductor-insulator-superconductor (SIS) process utilizing conventional AlO$_{\textrm{x}}$ as the insulating layer with a nominal critical current density of 1~kA~cm$^{-2}$. The fabrication is indicated in Fig.~\ref{fig:figure1} and uses electron-beam lithography and a chemical mechanical planarization (CMP). 

\begin{figure}[t]
\centerline{\includegraphics[width=0.78\columnwidth]{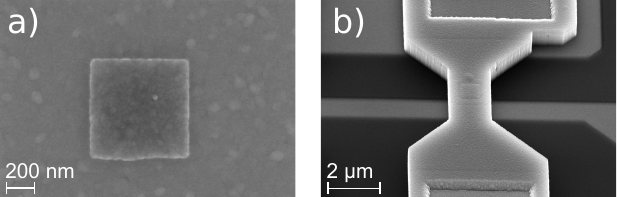}}
\caption{SEM image of a $(0.7\times 0.7)~\mu$m$^{2}$ AlO$_{\textrm{x}}$ junction. (a) After anodization. (b) With wiring electrode.}
\label{fig:figure2}
\end{figure}

The trilayer JJ is patterned using inductively coupled plasma reactive ion etching (ICP-RIE) for the Nb counter electrode (200~nm) and ion beam etching for the insulating junction layer ({20~nm Al + x~nm AlO$_{\textrm{x}}$}). This is followed by an anodization to electrically isolate the JJ edges and an SEM image of a $(0.7\times 0.7)~\mu$m$^{2}$ junction after this step is shown in Fig.~\ref{fig:figure2}(a). To pattern the Nb base electrode (160~nm), ICP-RIE is used once more. After depositing the SiO$_{2}$ insulation between the base and the Nb wiring, superfluous SiO$_{2}$ is removed via CMP to reveal the junction contacts and for planarization of the wafer surface. For the realization of superconducting connections between the base and wiring electrodes, vias in the insulation are opened by ICP-RIE. Subsequently, patterning of the resistance layer AuPd (75~nm) for the shunt resistors is done with a lift-off process and the final Nb wiring (560~nm) is structured using again ICP-RIE.  An SEM image of a $(0.7\times 0.7)~\mu$m$^{2}$ AlO$_{\textrm{x}}$ junction with the wiring electrode is shown in Fig.~\ref{fig:figure2}(b).

\subsection{Junction Characterization}

\begin{figure}[t]
\centerline{\includegraphics[width=0.88\columnwidth]{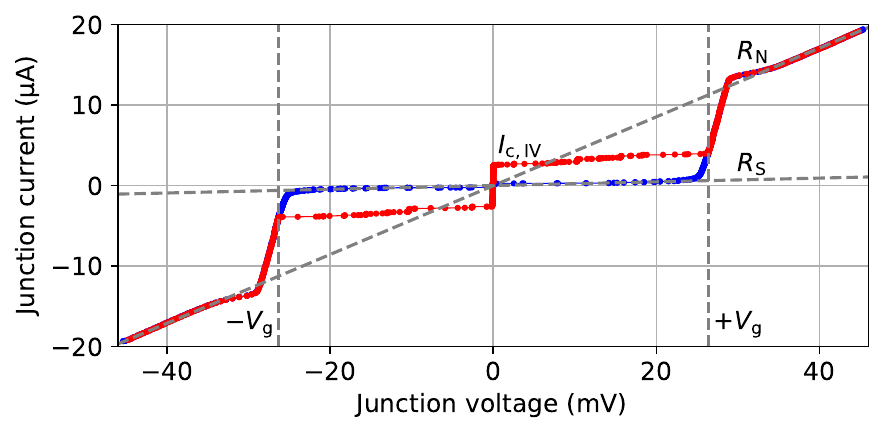}}
\caption{$I$-$V$ curve of the $(0.8\times 0.8)~\mu$m$^{2}$ junction array consisting of 10 JJs at 4.2~K. Division of the abscissa by 10 gives $V_{\textrm{g}}$ for a single junction.}
\label{fig:figure3}
\end{figure}

The $I$-$V$ curves of various series junction arrays were measured at 4.2~K. Exemplary data for the $(0.8\times 0.8)~\mu$m$^{2}$ JJ array are shown in Fig.~\ref{fig:figure3}, and the extracted parameters are listed in Table~\ref{tab:IVcurve}. The critical current $I_{\textrm{c,IV}}$ is significantly reduced compared to the nominal values and the results from the shunted JJs in the miniature SQUIDs (Section III) due to finite temperature and rf interference. The gap voltage $V_{\textrm{g}}$ is determined at $I_{\textrm{c,IV}}$ and the subgap resistance $R_{\textrm{S}}$ at 2~mV. $R_{\textrm{N}}$ is the normal resistance.

\begin{table}
\renewcommand{\arraystretch}{1.3}
\caption{Parameters of the submicron junctions at 4.2~K.}
\label{tab:IVcurve}
\centering
\begin{threeparttable}
{\begin{tabular}{c c c c c c c c}
\hline \hline
	JJ length	&	$I_{\textrm{c}}$&$I_{\textrm{c,IV}}$&		$V_{\textrm{g}}$	&			$R_{\textrm{N}}$&	 $R_{\textrm{S}}$ &	$R_{\textrm{S}}/R_{\textrm{N}}$	\\
	($\mu$m)	&		($\mu$A)			&			($\mu$A)			&	 	(mV)		&			($\Omega$)			& (k$\Omega$)				&								\\
\hline
		0.8			&		6.40$^{a}$		&					3.36			&		2.63			&				234				&			 4.79							&			20.5			\\
		0.7			&		4.90$^{a}$		&					2.34			& 	2.63			&				299				&			 6.01							&			20.0			\\
		0.6			&		3.60$^{a}$		&					1.64			&		2.63			&				397				&	 			9.98						&			25.1			\\
\hline
\end{tabular}}
\begin{tablenotes}[para,flushleft]
$^{a}$ nominal values

\end{tablenotes}
\end{threeparttable}
\end{table}

\section{SQUIDs based on sub-$\mu$m-sized JJs}

\subsection{SQUID Parameters}

For the evaluation of the Nb-AlO$_{\textrm{x}}$-Nb submicron-junction process, miniature SQUID magnetometers with $L_{\textrm{SQ}}=70~\textrm{pH}$ were fabricated. These have square junctions with side lengths of $0.8\,\mu$m and differ in their $R$. The design parameters for ${T=4.2}$~K are listed in Table~\ref{tab:Design_par} together with the values for a test SQUID with conventional JJs of 2.5~$\mu$m size. Current sensor SQUIDs with $L_{\textrm{SQ}}=80~\textrm{pH}$ and ${(0.7\times 0.7)}~\mu$m$^{2}$ JJs were also fabricated and characterized. All submicrometer-sized JJ devices were fabricated on a set of two wafers with nominally identical fabrication parameters.

\begin{table}[!t]
\renewcommand{\arraystretch}{1.3}
\caption{Parameters of miniature SQUIDs with $L_{\textrm{SQ}}=70~\textrm{pH}$ at 4.2~K.}
\label{tab:Design_par}
\centering
\begin{threeparttable}
\begin{tabular}{c c c c c c c c c c c}
\hline \hline
\#	&	JJ length	&$I_{\textrm{c}}$&		$C$		&	$R$							&	$\beta_{c}$&$	\beta_{L}$&$\varepsilon_{\textrm{w}}$&$\varepsilon_{\textrm{th}}$\\
		&	($\mu$m)	&			($\mu$A)	&($\mathrm{f\/F}$)&	($\Omega$)& 					&							&			($h$)								&			($h$)			\\
\hline
SQ-1	&			0.8			&		6.4			&			40$^{a}$			&					47		&			1.7			&			0.43		&				4.1$^{b}$					&		1.7		\\
SQ-2	&			0.8			&		6.4			&			40$^{a}$			&					34		&			0.90		&			0.43		&				5.3					&	2.4		\\
			&		2.5$^{c}$&		6.55		&				400					&					10		&			0.80		&			0.44		&				27				&	8.0		\\
\hline
\end{tabular}
\begin{tablenotes}[para,flushleft]
$^{a}$ estimated from 400~mK data\\
$^{b}$ mean value\\
$^{c}$ equivalent square length of octagonal shaped JJ
\end{tablenotes}
\end{threeparttable}
\end{table}

The $V$-$\Phi$ curve at 4.2~K of miniature SQUID SQ-2 is shown in the inset of Fig.~\ref{fig:figure4}. The indentation appearing at ${\approx90\,\mu\textrm{V}}$ corresponding to a Josephson frequency of ${\approx44~\textrm{GHz}}$ is caused by a $\lambda/4$ resonance of the interconnection stripline between SQUID and wire bonds. $I_{\textrm{c}}$ was calculated from the bias current needed for maximum voltage and we obtained a mean (eight devices) of $6.2\,\mu$A with a standard deviation of $0.5\,\mu$A for a single JJ. The capacitance of the ${(0.8\times 0.8)}~\mu$m$^{2}$-sized JJ was estimated from the $V$-$\Phi$ curve behavior of SQ-2 at 400~mK. Here, the measured $I_{\textrm{c}}$ increased from 6.6 to 7.2~$\mu$A in good agreement with the theoretical expectation~\cite{Ambegaokar1963a} resulting in a noise parameter ${\Gamma=2\pi k_{\textrm{B}}T/(I_{\textrm{c}}\Phi_{0})=2.3\times 10^{-3}}$. In this case, suppression of hysteresis by thermal noise is reduced and $\beta_{c}$ needs to be close to one. We estimate ${C\approx 40~\mathrm{f\/F}}$ leading to ${\beta_{c}=0.99}$ for SQ-2. This is also consistent with ${C\propto A}$ where $A$ is the JJ area. In contrast, SQ-1 ${(\beta_{c}=1.7)}$ showed hysteresis in the $V$-$\Phi$ curve at 400~mK.

\subsection{Performance}

\begin{figure}[t]
\centerline{\includegraphics[width=0.87\columnwidth]{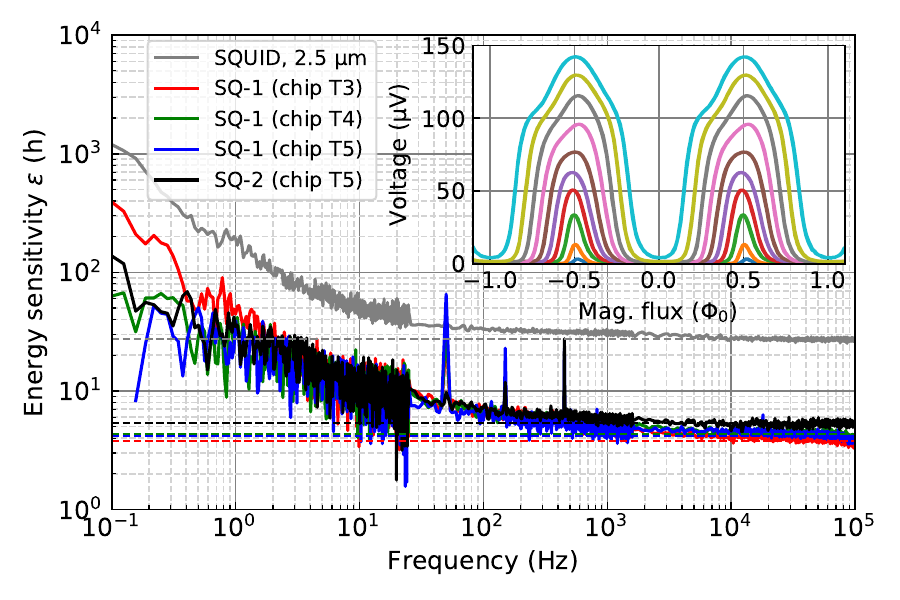}}
\caption{Energy sensitivity of four SQUIDs with $(0.8\times 0.8)~\mu$m$^{2}$ (SQ-1: $R=47\,\Omega$, SQ-2: $R=34\,\Omega$) and one SQUID with $(2.5\times 2.5)$~$\mu$m$^{2}$ JJs. The dashed lines give the white noise values $\varepsilon_{\textrm{w}}$. Inset shows $V$-$\Phi$ curves of SQ-2 with ${(0.8\times 0.8)}~\mu$m$^{2}$ JJs at 4.2~K for bias currents from 3 to 12 $\mu$A.}
\label{fig:figure4}
\end{figure}

We first discuss the performance determined at 4.2~K. The measurements were carried out in a liquid He bath using a direct read-out scheme and a single-stage configuration. The equivalent flux noise due to voltage and current noise of the XXF read-out electronics~\cite{magnicon} was subtracted to obtain the intrinsic SQUID flux noise $S_{\Phi}$. At 4.2~K our devices fulfill $\Gamma\beta_{L}< 0.2$ and the theoretical limit $\varepsilon_{\textrm{th}}$ can be estimated by
\begin{equation}
	\varepsilon_{\textrm{th}}\approx 2(1+\beta_{L})\Phi_{0}k_{\textrm{B}}T/I_{\textrm{c}} R
\label{eq:epsilon}
\end{equation}
which is valid for an arbitrary value of $\beta_{L}$~\cite{Clarke2004}.

As shown in Fig.~\ref{fig:figure4}, the intrinsic white energy sensitivity $\varepsilon_{\textrm{w}}$ was 4.1 and 5.3$\,h$ for SQ-1 and SQ-2, respectively. Those values are a factor of more than 5 better compared to our conventional technology with junction sizes of ${(2.5\times 2.5)}$~$\mu$m$^{2}$, which shows ${\varepsilon_{\textrm{w}}=27\,h}$. The slightly larger experimental $\varepsilon$ for SQ-2 is also expected theoretically using~(\ref{eq:epsilon}); however, the absolute values are about a factor of 2 larger. A possible origin might be suboptimal biasing or insufficient subtraction of the electronics contribution $S_{\Phi,\textrm{amp}}$ at 4.2~K as $V_{\Phi}$ is reduced in this case. A two-stage read-out scheme should be employed in future studies.

The energy sensitivity increases at lower frequencies with a typical value 37$\,h$ at 1 Hz corresponding to 900~n$\Phi_{0}$~Hz$^{-1/2}$, which was reproducible for the different devices. Interestingly, the low-frequency noise follows a $1/f^{\alpha}$ behavior with two distinct $\alpha$-values. For the $2.5~\mu$m-sized JJ SQUID and SQ-1 (chip T3), we find ${\alpha\sim1}$ consistent with critical current fluctuations as the origin. In contrast, the remaining SQUIDs show ${\alpha\sim0.6}$ at 4.2~K as has also been observed previously~\cite{Drung2011,Schmelz2017}.

\subsection{Temperature Dependence}
\begin{figure}[t]
\centerline{\includegraphics[width=0.87\columnwidth]{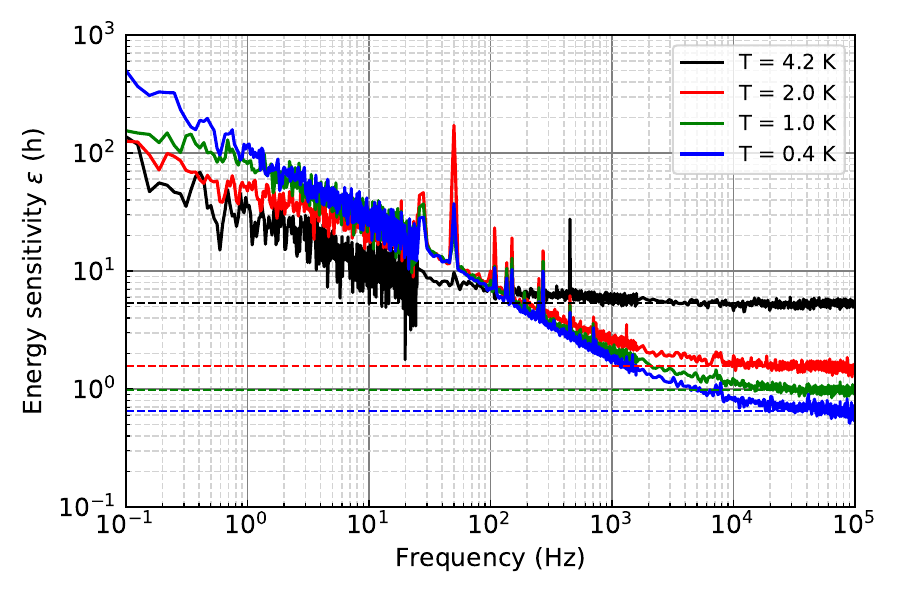}}
\caption{Energy sensitivity of SQ-2 with $(0.8\times 0.8)~\mu$m$^{2}$ for different temperatures. The dashed lines give the white noise values $\varepsilon_{\textrm{w}}$.}
\label{fig:figure5}
\end{figure}

Measurements below 4.2~K were carried out on a pulse-tube cooler equipped with a $^{3}$He sorption unit with the samples mounted in the vacuum space inside an aluminum and lead shield. No special precautions were made to thermally anchor the device to the cold stage of the cryostat. Consequently, the thermal link is mainly provided by the copper wiring used for biasing the SQUID.

Fig.~\ref{fig:figure5} shows the energy sensitivity for temperatures down to 400~mK together with the fitted white noise values $\varepsilon_{\textrm{w}}$ (dashed lines) for SQ-2. In agreement with our previous work~\cite{Drung2011}, we observe a reduction in the white noise level $\varepsilon_{\textrm{w}}$, an increase of the low-frequency noise and a shift of the $1/f$-noise corner to higher frequencies with decreasing temperatures. 

\begin{figure}[t]
\centerline{\includegraphics[width=0.87\columnwidth]{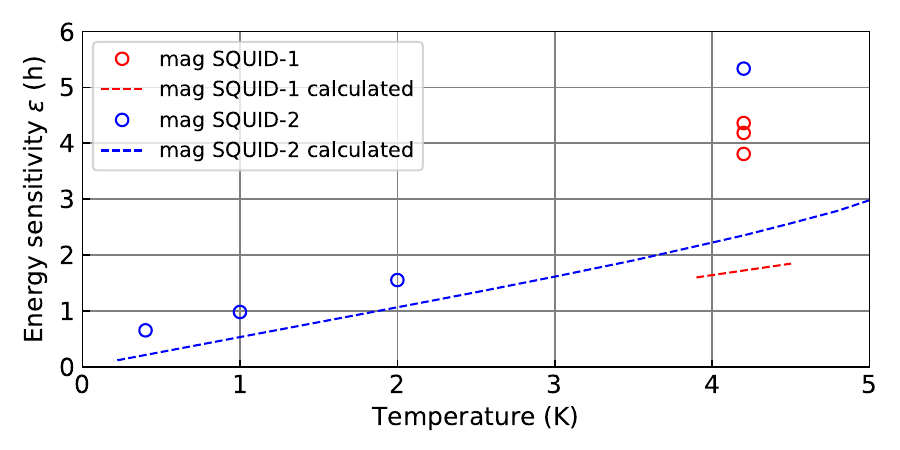}}
\caption{Energy sensitivity of magnetometer SQUIDs with $(0.8\times 0.8)~\mu$m$^{2}$ for different temperatures together with the theoretical curves.}
\label{fig:figure6}
\end{figure}

In discussing $\varepsilon_{\textrm{w}}(T)$, we refer to Fig.~\ref{fig:figure6}, which also contains the 4.2~K-values for all SQ-1 chips. Data for lower temperatures are not included due to the aforementioned hysteresis. At 400~mK, we determine $\varepsilon_{\textrm{w}}=0.65\,h$ for SQ-2. In addition, the theoretical curve obtained by combining~(\ref{eq:epsilon}) and the temperature dependence $I_{\textrm{c}}(T)$, as given in~\cite{Ambegaokar1963a} is shown. For ${T\leq 4.2}$~K, one finds ${I_{\textrm{c}}(T)/I_{\textrm{c}}(0)\approx 1}$ so that $\varepsilon_{\textrm{w}}(T)$ is dominated by the linear $k_{\textrm{B}}T$-term in~(\ref{eq:epsilon}). For the lower temperatures there is reasonable agreement, whereas at 4.2~K the measured $\varepsilon_{\textrm{w}}$ are significantly larger as already mentioned above.

It is worth discussing the temperature dependence of the low-frequency behavior in more detail. The unusual increase with decreasing temperature has been observed before~\cite{Wellstood1987} and for 4.2 and 0.4~K the excess noise follows roughly $1/f^{\alpha}$ with ${\alpha\sim0.6}$. However, for the intermediate temperatures of 1~K and 2~K, a low-frequency plateau emerges, which is particularly clear for the 1~K data. Whether this behavior is intrinsic to the fabrication technology remains unclear and requires further studies on a larger number of samples.

\subsection{Fully Integrated Current Sensor SQUIDs}

\begin{figure}[t]
\centerline{\includegraphics[width=0.87\columnwidth]{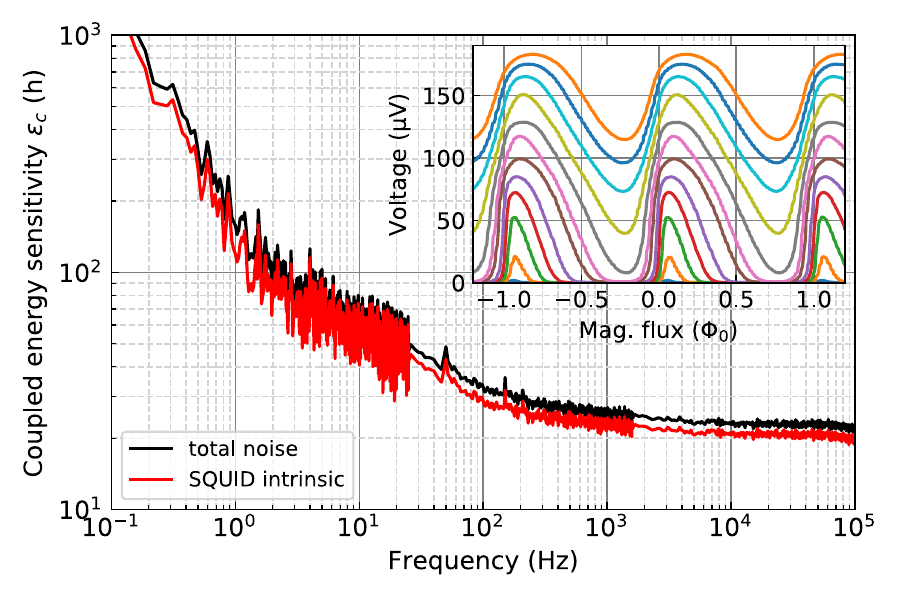}}
\caption{Coupled energy sensitivity of fully integrated current sensor SQUID with $(0.7\times 0.7)~\mu$m$^{2}$ JJs. Inset shows the set of $V$-$\Phi$ curves for bias currents ranging from 2, 4 to 14~$\mu$A.}
\label{fig:figure7}
\end{figure}

Based on these results, integrated current sensors were designed and fabricated having a SQUID inductance of 80~pH and $(0.7\times 0.7)~\mu$m$^{2}$ JJs. For the input circuit consisting of a 400~nH input coil, a proven double-transformer design was used with a coupling coefficient of ${k=0.75}$~\cite{Drung2007}. The expected white coupled energy sensitivity of this design is approximately 10$\,h$ at 4.2~K. In Fig.~\ref{fig:figure7}, the results show that we achieve an $\epsilon_{c,\textrm{w}}$ of 20$\,h$. The mismatch is most likely due to a parasitic resistance between SQUID and input coil of 75~$\Omega$. The low-frequency noise follows approximately $1/f$ and the set of $V$-$\Phi$ curves, given in the inset of Fig.~\ref{fig:figure7}, does not show any irregularities, e.g., due to resonances, for SQUID voltages up to $\sim 180~\mu$V. This could be avoided by the implementation of on-chip rf-filters compared to the miniature magnetometers.

Due to a limited number of fabrication runs the technology has not reached maturity, but the insufficient isolation observed in the current sensor device has been eliminated. Initially, as shown here, a good homogeneity of the JJs critical current could be achieved. However, after extensive alterations in the production technology, the restoration of this condition turns out to be difficult and is an ongoing task.

\section{Conclusions}
In this work, we presented a proof of concept of a new generation of SQUID sensors based on submicrometer-sized JJs. The fabrication technology was adapted from an established process for SNS junctions and is based on electron-beam lithography and CMP. Small test magnetometer SQUIDs with ${(0.8\times 0.8)}~\mu$m$^{2}$ JJs validated the approach for which we obtained energy sensitivities of about $5\,h$ at 4.2~K. When the devices were cooled to 400~mK, we achieved below $1\,h$ in reasonable agreement theoretical expectations. A current sensor SQUID utilizing a double transformer scheme suffered from an insufficient isolation and reached $\varepsilon_{\textrm{w}}$ of $20\,h$. For a reliable fabrication technology, we expect a performance of $10\,h$ at 4.2 K and about $1\,h$ when cooled to 400~mK. In order to minimize electronics noise contributions, a two-stage read-out scheme will be employed. 

Recent improvements in liquid He glass-fiber dewar construction circumvented the limiting thermal noise contributions, and one can expect a significant impact of such novel devices in biomagnetic measurements. In addition, they will be very useful in fundamental science experiments where SQUIDs are often operated in a superconducting shield environment and cooled down for increased sensitivity.

\section*{Acknoledgement}
The authors thank M. Petrich, J. Felgner, K. St\"orr and Th. Weimann for fabrication.

\bibliographystyle{IEEEtran}
\bibliography{TAS_2019_0303_R2}

\end{document}